\documentclass{csmagclass}														

\begin{document}		
\headings																															 %

\title{{Pairwise entanglement in double-tetrahedral chain with different Land\'e g-factors of the Ising and Heisenberg spins }}
\author{ L. G\'ALISOV\'A\thanks{Corresponding author: galisova.lucia@gmail.com}}
\affil{{Institute of Manufacturing Management, Faculty of Manufacturing Technologies with the seat in~Pre\v{s}ov,Technical University of Ko\v{s}ice, Bayerova~1, 080~01 Pre\v{s}ov, Slovakia}}

\maketitle

\begin{Abs}
The pairwise entanglement is exactly examined in the spin-1/2 Ising-Heisenberg double-tetrahedral chain with different Land\'e g-factors of the Ising and Heisenberg spins at zero and finite temperatures. It is shown that the phenomenon present in quantum non-chiral ground states is twice as strong as in quantum chiral ground states and that it gradually diminishes with increasing temperature until it completely vanishes at a certain threshold temperature. It is also demonstrated that the strong magnetic field maintains a weak thermal entanglement quite far from the saturation field, although the ground state is non-entangled.
\end{Abs}
\keyword{Ising-Heisenberg chain, thermal entanglement, chirality, exact results}
\section{Introduction}
Quantum entanglement belongs to the most sought phenomena in modern science mainly due to its extensive application potential, e.g. in quantum communication~\cite{Nie10} or quantum biology~\cite{Ball11}. In condensed-matter physics, this phenomenon provides a new perspective in understanding collective features of quantum many-body systems~\cite{Ami08}. 
Regarding to novel aspects of the quantum entanglement, quantum spin chains are of particular interest, since they provide a suitable playground for exact study of the phenomenon~\cite{Zad17}. 

In this paper, we will examine the quantum entanglement in the frustrated spin-$1/2$ Ising-Heisenberg double-tetrahedral chain with different Land\'e g-factors of the Ising and Heisenberg spins. The model was originally proposed and resolved by V.~Ohanyan {\it et al.}~\cite{Ant09,Oha10}, and recently re-examined in context of the magnetocaloric effect~\cite{Gal18}. Following the previous findings, the goal of this paper is to investigate how the spin chirality and the applied longitudinal magnetic field affect on an intensity of the pairwise entanglement in the model at zero and finite temperatures.

\section{Model and exact calculation of the concurrence}
To be specific, we consider the frustrated spin-$1/2$ Ising-Heisenberg double-tetrahedral chain of $N$ single Ising spins ($N\to\infty$), which regularly alternate with the XXZ-Heisenberg triangular clusters (see Fig.~1 in Ref.~\cite{Ant09} or~\cite{Gal18}). The  Hamiltonian of the model reads:
\begin{eqnarray}
\label{eq:H}
\hat{\cal H} = \sum_{i=1}^{N}\Big[J_H\sum_{j=1}^3\big(\hat{\mathbf S}_{i,j}\cdot\hat{\mathbf S}_{i,j+1}\big)_{\Delta}
 + J_I\sum_{j=1}^3\hat{S}_{i,j}^z(\hat{\sigma}_i^z+\hat{\sigma}_{i+1}^z) - h_H\sum_{j=1}^3\hat{S}_{i,j}^z - h_I\hat{\sigma}_i^z\Big].
\end{eqnarray}
In above, $\big(\hat{\mathbf S}_{i,j}\cdot\hat{\mathbf S}_{i,j+1}\big)_{\Delta}=\Delta(\hat{S}_{i,j}^x\hat{S}_{i,j+1}^x + \hat{S}_{i,j}^y\hat{S}_{i,j+1}^y)+\hat{S}_{i,j}^z\hat{S}_{i,j+1}^z$, where $\Delta$ is the exchange anisotropy parameter, and $\hat{\sigma}_i^z$, $\hat{S}_{i,j}^\alpha$ ($\alpha = x,y,z$) are the spatial components of the spin-$1/2$ operators related to the Ising spin at the $i$th nodal lattice site and the Heisenberg spin at the $j$th vertex of the adjacent (also $i$th) triangular cluster, respectively. For simplicity, the periodic boundary conditions $\hat{\sigma}_{N+1}^z=\hat{\sigma}_{1}^z$, $\hat{S}_{i,4}^\alpha = \hat{S}_{i,1}^\alpha$ are assumed. The parameter $J_H$ labels the anisotropic XXZ Heisenberg interaction between spins in triangular clusters, while $J_I$ marks the Ising interaction between nodal spins and spins from adjacent triangular clusters. Finally, the terms $h_{X} = g_{X}\mu_{B}h$ ($X=H$ or $I$) distinguish an action of the longitudinal magnetic field $h$ on the Heisenberg and Ising spins, which generally have different gyromagnetic factors $g_H$, $g_I$ ($\mu_B$ is Bohr magneton). 

As demonstrated in Refs.~\cite{Ant09,Gal18}, the considered quantum chain belongs to   lattice-statistical models with exact closed-form solution for the partition function and basic thermodynamic quantities. Some of them, namely the sublattice magnetization $M_{2} = \frac{1}{3N}\langle\sum_{j=1}^3\hat{S}_{i,j}^z\rangle$ and the pair correlation functions $C_{SS}^{(x,y)}=\langle\hat{S}_{i,j}^x\hat{S}_{i,j+1}^x\rangle = \langle\hat{S}_{i,j}^y\hat{S}_{i,j+1}^y\rangle$, $C_{SS}^{z}=\langle\hat{S}_{i,j}^z\hat{S}_{i,j+1}^z\rangle$, corresponding to the Heisenberg spins (for more computational details we refer the reader to the original work~\cite{Ant09}), can be directly utilized for a rigorous calculation of the physical quantity called concurrence~\cite{Woo98,Ami04}:
\begin{eqnarray}
\label{eq:C}
{\cal C} = 2{\rm max}\left\{0, 2\left|C_{SS}^{(x,y)}\right| - \sqrt{\bigg(\frac{1}{4} + C_{SS}^{z}\bigg)^{\!2} - M_{2}^2}\right\}.
\end{eqnarray} 
We note that the concurrence~(\ref{eq:C}) represents feasible measure of pairwise quantum entanglement of the Heisenberg spins from the same triangular claster at zero as well as finite temperatures.

\section{Numerical results and discussion}
In this section, we will discuss the quantum entanglement in the spin-$1/2$ Ising-Heisenberg tetrahedral chain with the easy-plane Heisenberg interaction (we set  $\Delta=2$ for simplicity) and the antiferromagnetic Ising interaction $J_I>0$. According to the previous results~\cite{Ant09,Oha10,Gal18}, this particular version of the model exhibits the most intriguing quantum features. For simplicity, we will consider the Land\'e g-factors $g_I=6$ and $g_H=2$ for the Ising and Heisenberg spins, which coincide with real gyromagnetic ratios for Co$^{2+}$ and Cu$^{2+}$ ions, respectively.

\begin{figure}[h!]
\includegraphics[width=8.5cm]{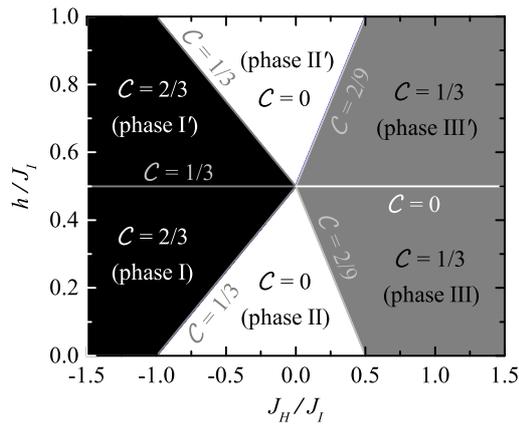}
\caption{The ground-state phase diagram in the $J_H/J_I-h/J_I$ plane for $J_{I}>0$ and $\Delta =2$ and $g_I=6$, $g_H=2$ supplemented with a density plot of the concurrence ${\cal C}$ in grey scale.}
\label{fig1}
\end{figure}

A diversity of the quantum entanglement in the ground state of the considered chain is obvious from Fig.~\ref{fig1}, where its ground-state phase diagram in the $J_H/J_I-h/J_I$ plane supplemented with a density plot of the quantity~(\ref{eq:C}) in grey scale is depicted. Clearly, three different values of the concurrence ${\cal C}$ may be observed in six diferent ground-state phases I, I$^{\prime}$, II, II$^{\prime}$, III, III$^{\prime}$. In our notation, the primed/unprimed phases are characterized by the parallel/antiparallel orientation of the nodal Ising spins with respect to the magnetic field direction.
The highest value ${\cal C}=2/3$, which can be found in the phases I, I$^{\prime}$ for the ferromagnetic Heisenberg coupling $J_H<0$ at the magnetic fields $0.5 + 0.5J_H/J_I<h/J_I<0.5$, $0.5<h/J_I<0.5 - 0.5J_H/J_I$, indicates the strongest pairwise entanglement in the Heisenberg triangles. Both the phases are unique quantum ground states, where the Heisenberg spins from each triangular cluster are in a symmetric quantum superposition of three possible up-up-down spin states. On the other hand, ${\cal C}=1/3$ observed in the phases III and III$^\prime$ when $J_H>0$ and $0.5 - J_H/J_I<h/J_I<0.5$ and/or $0.5<h/J_I<0.5 + J_H/J_I$ points to a half weaker quantum entanglement between any two Heisenberg spins from the same triangular cluster due to two possible chiral degrees of freedom of each triangle. Finally, the zero concurrence ${\cal C}=0$ indicates a non-entangled spin arrangement of the Heisenberg spins in the remaining phases II, II$^\prime$. Interestingly, ${\cal C}$ is zero also at each point of the phase boundary ${\rm III}-{\rm III}^{\prime}$ due to a higly non-trivial macroscopic degeneracy of the model. 
By contrast, the chain remains partially entangled along all other ground-state boundaries, as confirmed by the finite value of the concurrence ${\cal C}=1/3$ observed along the boundaries ${\rm I}-{\rm I}^{\prime}$, ${\rm I}-{\rm II}$, ${\rm I}^{\prime}-{\rm II}^{\prime}$ and ${\cal C}=2/9$ found along the boundaries ${\rm II}-{\rm III}$, ${\rm II}^{\prime}-{\rm III}^{\prime}$ (see Fig.~\ref{fig1}).  For more details on spin ordering and degeneracies of the individual ground-state phases and phase boundaries see Refs.~\cite{Ant09,Oha10,Gal18}.

\begin{figure}[h!]
\includegraphics[width=8.5cm]{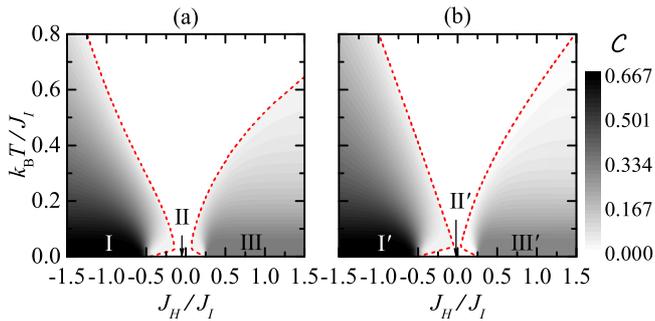}
\caption{Density plot of the concurrence ${\cal C}$ along with the threshold temperature $k_{\rm B}T_{th}/J_I$ (red dashed lines) in the  $J_H/J_I-k_{\rm B}T/J_I$ plane for  (a)~$h/J_I=0.25$ and (b)~$h/J_I=0.75$.}
\label{fig2}
\end{figure}

\begin{figure}[h!]
\includegraphics[width=8.5cm]{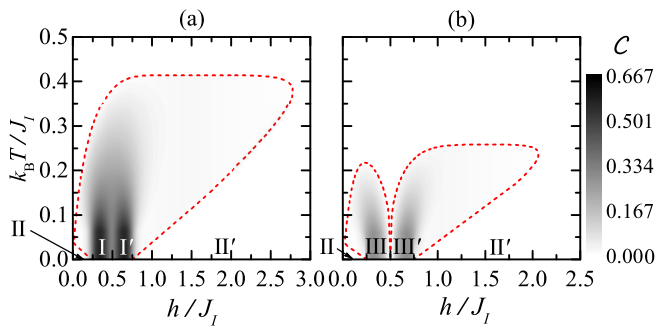}
\caption{Density plot of the concurrence ${\cal C}$ along with the threshold temperature $k_{\rm B}T_{th}/J_I$ (red dashed lines) in the $h/J_I-k_{\rm B}T/J_I$ plane for  (a)~$J_H/J_I=-0.5$ and (b)~$J_H/J_I=0.25$.}
\label{fig3}
\end{figure}

The effect of the temperature on pairwise entanglement in the model can be well understood from Figs.~\ref{fig2} and~\ref{fig3}, which display the grey-scale density plots of the concurrence ${\cal C}$ and the threshold temperature $k_{\rm B}T_{th}/J_I$ (red dashed lines) in the parameter planes $J_H/J_I-k_{\rm B}T/J_I$ and $h/J_I-k_{\rm B}T/J_I$, respectively. The threshold temperature $k_{\rm B}T_{th}/J_I$ has been numerically obtained from Eq.~(\ref{eq:C}) by setting ${\cal C}=0$. In both the figures, $k_{\rm B}T_{th}/J_I$ unambiguously delimites the quantum entanglement at finite temperatures and also in the asymptotic limit $k_{\rm B}T/J_I\to 0$. In agreement with the ground-state analysis, $k_{\rm B}T_{th}/J_I$ drops down to zero at the critical values of $J_H/J_I$ and $h/J_I$, which correspond to the ground-state phase boundaries between quantum and classical phases and the phase boundary ${\rm III}-{\rm III}^{\prime}$.
Moreover, the concurrence ${\cal C}$ observed in the phases I, I$^{\prime}$, III and/or III$^{\prime}$ gradually declines with increasing temperature until it completely vanishes at a certain threshold temperature. The stronger the Heisenberg interaction $J_H$ is, the higher $k_{\rm B}T_{th}/J_I$ can be observed, regardless of the sign of $J_H$ (see Fig.~\ref{fig2}). In addition, the pairwise entanglement of the Heisenberg spins may appear at finite temperatures even if the ground state is non-entangled. This happends when the temperature increase supports a creation of quantum excited spin states in the system. 
Consequently, two threshold temperatures can be observed around the ground-state boundaries separating the entangled phase from the non-entagled one (see Figs.~\ref{fig2} and~\ref{fig3}). It is noteworthy that above the saturation fields $h_{s1}/J_I = 0.5 - 0.5J_H/J_I$ ($J_H<0$), $h_{s2}/J_I = 0.5 + J_H/J_I$ ($J_H>0$), which correspond to the phase transitions ${\rm I}^{\prime}-{\rm II}^{\prime}$, ${\rm III}^{\prime}-{\rm II}^{\prime}$, respectively, the lower threshold temperature monotonically increases as the magnetic field is lifted from its saturation value, while the higher one remains constant from a certain value $h>h_{s1(2)}$. Both the threshold temperatures merge quite far from saturation fields (see Fig.~\ref{fig3}). One can thus conclude that the strong magnetic field quite well maintains a very weak pair entanglement in the studied chain at relatively high temperatures.

\section{Summary}
This work deals with the quantum pairwise entanglement in the exacty solvable frustrated spin-1/2 Ising-Heisenberg double-tetrahedral chain with different Land\'e g-factors of the Ising and Heisenberg spins. As has been shown, the phenomenon observed in ground states I, I$^{\prime}$ with the symmetric quantum superposition of three possible spin states of the Heisenberg trimers is twice as strong as in the macroscopically degenerated chiral ground states III, III$^{\prime}$. Generally, the quantum entanglement found in the quantum ground states (both the non-chiral and chiral ones) gradually diminishes with the increasing temperature until it completely vanishes at a certain threshold temperature. Moreover, the pairwise entanglement may emerge at finite temperatures, although the ground state is non-entangled, and that the strong magnetic field maintains a weak thermal entanglement quite far from the saturation field.

To conclude, the investigated spin-1/2 Ising-Heisenberg double-tetrahedral chain can be very easily extended to planar system of various topologies, as it belongs to a class of bond-decorated lattice-statistical models. Our future investigations will be focused in this direction.

\section{Acknowledgement}
This work was financially supported by the grant of Slovak Research and Development Agency under the contract No.~APVV-16-0186.
%
%



\begin{thebibliography}{99}\setlength{\itemsep}{-0.10cm}
\bibitem{Nie10} 
M.~A. Nielsen and I.~L. Chuang, \textit{Quantum Computation
and Quantum Information}, DOI:~10.1017/CBO9780511976667 Cambridge University Press, Cambridge 2010.
\bibitem{Ball11}
P. Ball, \textit{Nature} {\bf 474}, 272 (2011). DOI:~10.1038/474272a
\bibitem{Ami08}
L. Amico, R. Fazio, A. Osterloh, and V. Vedral, \textit{Rev. Mod. Phys.} {\bf 80}, 517 (2008).  DOI:~10.1103/RevModPhys.80.517
\bibitem{Zad17}
H. A. Zad, H. Movahhedian, \textit{Int. J. Mod. Phys. B} {\bf 31}, 1750094 (2017). 
DOI:~10.1142/S0217979217500941
\bibitem{Ant09}
D. Antonosyan, S. Bellucci, and V. Ohanyan, \textit{Phys. Rev. B} {\bf 79}, 014432 (2009). 
DOI:~10.1103/PhysRevB.79.014432
\bibitem{Oha10}
V. Ohanyan, \textit{Phys. Atom. Nucl.} {\bf 73}, 494 (2010). DOI:~10.1134/S1063778810030129
\bibitem{Gal18}
L. G\'alisov\'a, D. Kne\v{z}o, \textit{Phys. Lett. A} {\bf 382}, 2839 (2018). 
DOI:~10.1016/j.physleta.2018.06.012
\bibitem{Woo98}
W. K. Wooters, {\it Phys. Rev. Lett.} {\bf  80}, 2245 (1998).
DOI:~10.1103/PhysRevLett.80.2245
\bibitem{Ami04}
L. Amico, A. Osterloh, F. Plastina, R. Fazio, and G.M. Palma, {\it Phys. Rev. A} {\bf  69}, 022304 (2004).
DOI:~10.1103/PhysRevA.69.022304
\end{thebibliography}
\end{document}